\documentclass[pdflatex,sn-mathphys]{sn-jnl}

\jyear{2022}%

\theoremstyle{thmstyleone}%
\theoremstyle{thmstyletwo}%

\theoremstyle{thmstylethree}%

\begin{document}

\title{Evaluation of User Perception on Biometric Fingerprint System}


\author*[1]{\fnm{Jones} \sur{Yeboah}}\email{yeboahjs@mail.uc.edu}

\author[1]{\fnm{Victor} \sur{Adewopo}}\email{Adewopva@mail.uc.edu}
\equalcont{These authors contributed equally to this work.}

\author[1]{\fnm{Sylvia} \sur{Azumah}}\email{azumahsw@mail.uc.edu}
\equalcont{These authors contributed equally to this work.}
\author[1]{\fnm{Izunna} \sur{Okpala}}\email{okpalaiu@mail.uc.edu}
\equalcont{These authors contributed equally to this work.}

\affil*[1]{\orgdiv{School of Information Technology}, \orgname{University of Cincinnati}, \orgaddress{\street{2600 Clifton Avenue}, \city{Cincinnati}, \postcode{45221}, \state{Ohio}, \country{USA}}}

\begin{huge}
\textbf{Springer Copyright Notice}
\end{huge}

\vspace{5mm} 

\begin{large}
Copyright(c) 2022
\end{large}
\vspace{5mm} 
\begin{large}

This work is subject to copyright. All rights are reserved by the Publisher, whether the whole or part of the material is concerned, specifically the rights of translation, reprinting, reuse of illustrations, recitation,broadcasting, reproduction on microfilms or in any other physical way, and transmission or information storage and retrieval, electronic adaptation, computer software,or by similar or dissimilar methodology now known or hereafter developed.\\

\textbf{Accepted to be published in: }The 2022 International Conference on Security and Management (SAM'22)
Las Vegas, USA, July 25 - 28; https://sam.udmercy.edu/sam22/
\end{large}

\vspace{5mm} 

\abstract{
Biometric systems involve security assurance to make our system highly secured and robust. Nowadays, biometric technology has been fixed into new systems with the aim of enforcing strong privacy and security. Several innovative system have been introduced, and most of them have biometrics installed to protect military bases, banking machines, and other sophisticated systems, such as online tracking systems. Businesses can now focus on their core functions and feel confident about their data security. Despite the benefits and enhancements in security that biometrics offer, there are also some vulnerabilities. This study aimed to investigate the biometric vulnerabilities in a healthcare facility and propose possible countermeasures for biometric system vulnerabilities.}





\maketitle
\section{Introduction}
A biometric device is installed in many systems, ranging from simple mobile application systems to sophisticated organization systems. In general, biometric systems are designed to identify and verify an individual's identity through the use of their physical characteristics (including their face, iris, fingerprint, DNA, or hand geometry) and behavioral characteristics (including their speech, gait, or signature) \cite{al2019automated}. It also offers a reliable identity management system that can prevent identity theft and restrict access to various resources. An important aspect of any identity management system is the identification of the individual accurately \cite{jain2011introduction}. 
The biometric behaviors ensure the connection between a person and their unique identity since the process requires user authentication. Although most systems proved that they are reliable, there are gaps in the security of some biometric systems, and more effort is required for quality service to be provided.
This work presents the vulnerabilities and the enormous benefits that come with new state-of-the-art security technologies.
Khushk et al. \cite{iqbal2005overview} explain that  “biometrics can be defined as an automated method of verifying or recognizing the identity of an individual person based on a physiological or behavioral characteristic. There can be changes to passwords easily, but a biometric identification of an individual is typically permanent~\cite{jain2006biometrics}. 

Security of biometrics is very crucial and the application of biometric technologies is more prevalent in three main areas: smartphones, wearables, and online shopping. Smartphones have biometrics built in to increase the security and safety of the data by using fingerprint recognition, voice recognition, and face recognition. Using wearable devices, it is possible to identify the behavioral and biological characteristics of individuals. It is now possible to secure online systems by deploying biometrics for access control within online systems by utilizing fingerprints, iris, and facial recognition instead of credentials \cite{tiwari2015review,sabhanayagam2018comprehensive}.   
Biometric systems are designed to ensure that only genuine users can authenticate, while impostor users cannot ~\cite{jain2012biometric} which also introduces the privacy requirements on biometrics, due to the particular nature of biometrics carrying much information about the person that cannot be easily revoked when compromised. It is important to emphasize that security and privacy requirements are not the same, and they need a different approach.
The following research questions were postulated to address the strengths and weaknesses of biometric vulnerabilities.
\begin{itemize}
  \item \textbf{RQ1:} \textit{What are the critical vulnerability issues in biometrics?}
  \item \textbf{RQ2:} \textit{What are the solutions required to mitigate the vulnerabilities detected in biometrics?}
  \item \textbf{RQ3:} \textit{How can we evaluate issues related to biometric system vulnerabilities?}
\end{itemize}

\section{Background Literature}

The concept of biometric technology stems from the need for originality in human verification. This creates an extra layer of information security. According to an article by Clodfelter in 2010 \cite{CLODFELTER2010181}, the benefit of biometric technology goes beyond security. It cuts across security and decision making \cite{CLODFELTER2010181}. Data captured through a biometric system can be used to make informed decisions in an organization \cite{udebuanadecision,CLODFELTER2010181}. One of the earliest biometric technology that has been lumped together under the umbrella of digital forensics is the fingerprint recognition system \cite{rashid2008security}. With the proliferation of a myriad of tech devices, the need for better security has emerged, coupled with the fact that the usage of the data from many sources has become a lightning rod for privacy and human rights concerns \cite{bertino2016data}.


Based on the rising concerns for security, different technologies are being developed and deployed to secure and ease the concerns of organizations and users. Some of these technologies include smart cards, antivirus software, biometrics, firewalls, password-protected accounts, and intrusion detection/prevention systems, to name a few.  
These rising concerns about security also move technology away from people in terms of “ease of use” of operations. Thus, there is a need to integrate the required functionality of authentication and control to safeguard the data and information and “ease of use”. 
According to Chellappa, Wilson, and Sirohey  \cite{chellappa1995human}, biometric technology is one of the innovations that is being claimed to achieve this objective. Khushk and Iqbal  \cite{iqbal2005overview} explains that “biometrics can be defined as an automated method of verifying or recognizing the identity of a living person based upon a physiological or behavioral characteristic; that is, it's based upon something we are or something we do”.   Unlike passwords, biometric identification of an individual for the most part is permanent and cannot be easily changed.  
Several applications have found a natural fit in biometric deployments and that is why biometric security is of paramount importance. The public acceptance of biometric systems depends on the system's identity management and authentication mechanisms that ensure the security of the modules and at the same time, protects the privacy of the users.
In biometric systems, security usually revolves around assuring that only real users can authenticate and the templates stored in the database will also be protected against unauthorized access. The last statement also introduces the privacy requirements of biometrics, due in part to the fact that biometrics carry much information about the person. This data cannot be easily retrieved if compromised, it is important to note that security and privacy requirements are two different things. An accurate system must minimize both False Acceptance Rate and False Rejection Rate, thus achieving the maximum performance possible \cite{kalunga2016development}.


\subsection{Biometric Traits}

The distinguishing quality or characteristics of an individual embedded in a biometric system known as biometric traits can be divided into the behavioral and physiological traits \cite{hossain2011human}. The behavioral traits encompass the basic human functions as it engages in different activities while the physiological traits involve the physical characteristics of an individual \cite{vielhauer2005biometric,hossain2011human}.
Faundez-Zanuy et. al \cite{faundez2006biometric} explains that a good biometric trait must accomplish the following set of properties such as 1)
Universality: Every person should have the characteristics. 2) Distinctiveness: This refers to as uniqueness \cite{yun2002123}. The distinctive character of any two individuals should be enough to distinguish them from each other. 3) Permanence: The characteristic should be stable enough (concerning the matching criterion) overtime, and under different environmental conditions. 4) Collectability: Characteristics should be acquirable and quantitatively measured. 5) Acceptability: People should be willing to accept the biometric system. 6) Performance: There is identification accuracy, and the required time for a successful recognition must be reasonably good. 7) Circumvention: This is the ability of fraudulent people to fool the biometric system and make it negligible. Faundez-Zanuy et. al \cite{faundez2006biometric} identified that biometric traits can be split into two main categories: Physiological Biometrics: It is based on direct measurements of a part of the human body like a fingerprint. Behavioral Biometrics: It indirectly measures some characteristics of the human body like signature and key stroking recognition.  

\subsection{Types of Authenticators}

O'Gorman et al \cite{o2003comparing} classified authenticators into three distinctive categories: 1)Knowledge-Based Authenticators: It is characterized by obscurity or secrecy. An example of this type is the memorized password. This type can also include information that is not a secret (obscure). This can be loosely defined as an insecure password for most people. Using lastname or common words is a good use case in this category. In this case, a security drawback of a passcode is that each time it is shared for authentication, it becomes less secure \cite{o2003comparing}. 2) Object-Based Authenticators: It's characterized by physical possession. Physical keys are sometimes called metal keys. The main security drawback of a metal key is that once misplaced, it permits random intruders who find such keys to access the house. This is why many digital tokens also combine another factor, i.e. an associated password, to protect a stolen or lost token. There is a different advantage of a physical object used as an authenticator; in case it has been lost, the owner sees evidence of this and can act accordingly \cite{o2003comparing}. 3)ID-Based Authenticators: These are also characterized by uniqueness to one person. A passport, university diploma, credit card, driver’s license, etc., all belong in this category. So does a biometric, such as a voiceprint, eye scan, fingerprint, or signature. For both ID documents and biometrics, the central security defense is that they are difficult to copy or forge \cite{o2003comparing}. 

\section{Issues and Challenges}
In selecting a specific biometric technology, the potential challenges that should be considered include user group size, place of use, nature of use, ease of use, user training required, error incidence (age, environment, and health condition), security requirement needed, user acceptance level, long term stability including technology maturity and the cost associated \cite{yun2002123}.

\subsection{Vulnerabilities On Biometrics}
Abdulmonamm et al \cite{alaswad2014vulnerabilities}, noted that biometric attacks fall under these three (3) attacks. 1) Processing and Transmission Attacks: System processing is done locally or remotely, it is crucial for such transmission to be fully secured to avoid external attacks from intercepting or reading transmission. Encryption is mostly enforced on such systems to guarantee full encryption. However, some devices can prove otherwise. It is imperative for deployers to understand the data risk when exposed on transit. 2) Input Level Attacks: This attack is perpetrated to affect system performance and effectiveness. The most common vulnerability is spoofing, but bypassing has also been used in some cases. Overloading can also occur when a user attempts to bypass a system by overpowering it with the objective of generating errors or breaking down the input device. However, some system failures can also be attributed to human error \cite{alaswad2014vulnerabilities}. 3) Back-end Attacks: The main goal of the attacker is to have unauthorized control of storage databases. Back-end attack is mostly caused by targeting the template storage database. An attacker can introduce themselves into the system without following the laydown procedure for enrollment if they can discover ways of introducing templates directly into the storage database \cite{jain2015attacks}. A criminal offense is committed when a technical device is used to gain unauthorized access to a system. The Electronic Communications Privacy Act (ECPA) was passed to address such offenses enacted to protect electronic communications in storage and transit \cite{adewopo2021exploring,adewopo2022deep}. 

\subsection{Mitigating the vulnerabilities of biometrics}
Abdulmonam et al \cite{alaswad2014vulnerabilities}, propose a technique to mitigate processing and transmission level attacks by identifying and using several encryption techniques that are important aspects of biometric security. A multi-factor authentication system was introduced that uses biometrics as well as smart cards and PINs or multiple biometrics to reduce the possibility of unauthorized access. The designers of biometric technologies should explore random or cued challenges when time permits. This is to ensure that there is a very strong authentication method to ensure the submission of genuine data by introducing multiple authentication methods \cite{alaswad2014vulnerabilities}. 

To mitigate input level attacks, systems must be built robustly and their basic functions should not malfunction even if they are overwhelmed. An enforced fallback process needs to be in place when a system is not functioning properly and backup measures are necessary to ensure that security systems can recover from failures \cite{alaswad2014vulnerabilities}. Approximately 70\% percent of IoT devices can be hacked and securing these devices is important as they improve the quality of life of people through the advancement of technology \cite{Sylvia}. Another method to mitigate back-end attacks involves enforcing data integrity, constraints, and other security features that can make it difficult for an intruder to get into the system. DDoS attacks can be thwarted with traffic monitoring and traffic analysis. 

\section{Methods}
We collected data from stakeholders (Staff and Patients) using the survey method of data collection. Survey and interview seem to be one of the most appropriate methodologies discussed in \cite{cirqueira2020scenario} to scope information about a design. Considering the research is presenting new ideas on biometrics usage in Ghana, the study was evaluated as a descriptive study and after careful consideration, we determined that qualitative analysis is the best approach for gaining a comprehensive understanding of the biometric system.

\subsubsection{Research Design}
This work check how biometric system adoption is being used by customers and staff at Acacia Medical Centre located in the capital city of Ghana, Accra. The center we chose for our study is able to provide us with an in-depth analysis of the challenges associated with the use of biometric systems. We randomly selected staff and patients of the medical center by sampling 20 staff members and 100 patients. Data collection procedures were observed during interviews with biometric company representatives, hospital administrators, IT security personnel, and biometric technical personnel. The participants' consent was sought before video recording the data during the interviews for further analysis. 
\subsection{Empirical Findings of Interview}
We interviewed representatives of the biometric systems providers and experts to learn more about the state of biometric systems in the healthcare industry and to gain a better understanding of the factors influencing the adoption and deployment of biometric systems in this field. As part of our research, we selected five interviewees to share their experiences, opinions, and domain knowledge about the implementation of biometrics in the Ghanaian healthcare sector. The selection was made based on our aim to interview different stakeholder representatives, who could provide us with information about different perspectives on biometrics in healthcare. Among the themes that emerged from the interview were users' perspectives on the current picture of biometric technology adoption and the security of biometric systems in the healthcare industry. 

\subsection{Analysis}
\textbf{Interview Analysis}
When implementing biometrics, the interviewee stressed the importance of three attributes: cost, privacy, and security. The participants believe physical biometrics are more secure and perform better when these three factors are taken into consideration, but depending on system performance, costs may be higher. Another interviewee said biometrics could improve security and privacy in healthcare. Additionally, it ensures the patient's identity and removes duplicate data. In response, organizations are demanding strong authentication and focusing on biometrics for the benefit of staff, and healthcare providers. As the interviewee explained the cost of the biometric solution versus the card system, a fingerprint and smart card combination was suggested for securing fingerprint devices and other mechanisms for data protection. According to a third interviewee, biometrics needs to be implemented. These devices are easy to use and extremely secure as users carry identification cards around with them. Further, healthcare organizations have access to a lot of sensitive patient data, so it is imperative that this data is protected. Therefore, there are many biometric authentication devices available on the market that can secure data in the most efficient way.
The fourth interviewee discussed how she views biometrics in healthcare. If they are not properly identified, elderly people can still be identified when they are sick and need acute care. In the ambulance, emergency room, and at home, biometrics can be used to establish patients' identities. Biometrics are more secure than passwords when dealing with medical records. Using biometrics to secure the entrances to special buildings and rooms for both medical staff and patients is most feasible for entry clearance points.  
In a fifth interviewee's workplace, physical keys are still used at certain locations, but they also use plastic cards with magnetic stripes and PIN codes. We need a new method of authentication. Currently, we are focused on all the national initiatives in this area, so the "new" techniques are coming, but they don't involve biometrics. This could be because we do not yet understand the biometric technique and the user community needs to be aware.

\subsubsection{BIOMETRIC PERFORMANCE MEASURES}
In order to examine and compare the performance of biometric technologies, there are four (4) key measures identified below that are usually used to test such systems. 1)False Acceptance Rate (FAR): Fault acceptance rates are also called Type I errors. FAR indicates the percent of impostors whose accounts are incorrectly approved as genuine \cite{conrad2017chapter}. This number should be as low as possible because most biometric systems aim to attain accurate identity authentication. 2)False Rejection Rate (FRR): The FRR is also known as “Type II error”. FRR is a measure of the percentage of genuine users that are incorrectly rejected. This number should be kept as low as possible to avoid causing inconvenience to the legitimate user\cite{Eberz,kalunga2016development}. As a general rule, this error is more acceptable since a second attempt can be made. 3)Equal Error Rate (EER), FAR, and FRR are related: A stringent requirement for FAR (as low as possible) will inadvertently increase the FRR. The point where the FRR is equal to FAR is given by this measure. Lowering the rate of EER will increase the performance of the system as it indicates a good balance in the sensitivity of the system\cite{Salloum}. 4)Crossover Error Rate (CER): The CER is a metric used to compare biometric technologies. Basically, this is the error rate at which FRR equals FAR. FRRs and FERs are used to determine biometric accuracy and the system's capability of allowing limited access to authorized users \cite{Sivaram}. The measures can differ widely depending on how sensitive the mechanism is set up to match the biometric. In such cases, it could be necessary to measure the hand geometry more precisely and the user's template more sensitively (increase sensitivity). In addition to reducing false-acceptance rates, false-rejection rates can also be increased. This makes it essential to understand how vendors calculate FARs and FRRs \cite{kalunga2016development, Sivaram}. As a result of their interdependence, plotting them against each other provides meaningful insights. Several types of sensitivity settings are shown on the plot of a hypothetical system. With a lower CER, accuracy is higher. It is also noteworthy that biometrics performed physically are usually more accurate than behavioral.

\subsection{Results}
The results and discussions are based on the collated data from the patients and staff at the Acacia Medical Centre. About one hundred and twenty questionnaires were filled out by staff and patients of Acacia Medical Centre, Acacia Health Insurance, and Technology departments.
We have approximately sixty percent (60\%) of male participants, while we have forty percent (40\%) of female participants. 30\% of the participants are between the ages of 20 and 30 years. There were 60\% of employees between 31 and 40 years old, 10\% between 41 and 45 years old, and no information was available on those over 46 years old. 
Approximately 60\% of our participants have obtained a graduate degree and above 70\% indicated they have four to six years of work experience which indicates that most participants are highly educated and experienced. 
Based on our study, we found staff members to be enthusiastic about adopting a biometric system due to its enormous benefits. According to this study, 14.5\% of the participants identified that Acacia Medical Centre had been the victim of enrollment attacks. In addition, ten (10) respondents representing 14.5\% of the sample population stated that they experienced a back-end attack. Input level attacks have also been reported by fifteen (15) respondents, representing 22 \%. Among the respondents, 18 participants experienced processing and transmission-level attacks and 23\% experienced other security issues.

 The study identified areas like 24/7 Availability of Biometrics Services, Information Security, and Efficient and Effective Customer Support as critical to the effective use of Biometrics. Except for internal and external customer education, where 22 percent are represented, we found a high percentage of 26 percent for all of the critical areas. In our study, we found that even though there has been a significant impact after the introduction of biometrics, staff were able to point out certain obstacles they encounter when using the biometric system. There were extremely high implementation costs, comprising 36\%, while infrastructure was lacking. Additionally, 18\% of the participants expressed security concerns, while only 10\% were concerned about the ever-changing technology trend. When patients face any challenges in using the biometric system, they are more likely to report them directly to the health center rather than using other options. There was also an increased level of satisfaction: 60 \% gave good reviews while 40 \% were unimpressed. It was observed that most patients were willing to recommend biometrics to friends.

\subsection{Discussion}
This study was conducted to investigate the interest of Ghanaians in biometrics and was based on a case study of the medical center.  Acacia Medical Centre was selected for our study based on the availability of biometrics system infrastructure in the healthcare facility. The key study areas focused on identifying critical vulnerabilities for biometric systems and providing solutions.

\textbf{Vulnerable points: }
The first phase involves getting the client's biometric information. Databases are created during enrollment that has an invariant format that identifies distinct individuals. This format is retrieved from the database and compared to the new template to verify the client. The process is similar to setting up a password as users need to create a password and then enter their previous password to gain access.

\textbf{Attacks: }
In E-authentication, biometrics are used to bind an individual to their identity. There were several potential vulnerabilities identified, including fictitious data, conspiracy with employees, impersonation, and sales of PII in deep web forums \cite{vic}. As part of Acacia Medical Centre's commitment to modernizing the hospital's services, this study demonstrates Acacia Medical Centre's readiness to continuously improve it is biometric services. As a result of its enormous benefits, the medical center adopted biometric technology. Authentication with fingerprint scanners is reliable, whereas Iris and Retina scans are more secure since it requires copying an individual's retinal pattern. In 2018, Search Security stated that keystroke dynamics could be used to authenticate users based on the number of words they type \cite{qvintus2018evaluating}. 
In addition to illustrating some of the challenges that healthcare facilities are facing in Ghana, the study also revealed some of the resistance they experience to embracing technology-driven innovation. Biometric systems have a variety of benefits, such as reliability, cost-effectiveness, high-level identification, ease of use, scalability, and versatility. In part, illiteracy and lack of interest in technological advancement are responsible for higher execution costs. Our research revealed that Acacia medical center was very concerned about security and had put in place a variety of security measures. Only a small proportion of customers experienced biometric security problems as a result. 
Moreover, our study showed that the support services and contingency plans implemented in case of a crisis helped create trust among patients through a rapid response when issues arise, which led to more than half of the clients expressing their intention to tell their friends about the advantages of their biometric system.

\section{Conclusion}
Biometrics offer solutions and an effective way to the establishment  of who a person is. The purpose of this scheme is to make it difficult for perpetrators to cause damage to a rightful user by verifying a user's behavior. The best way to resolve biometric issues is through administrative channels to prevent a public uproar.
Biometrics are not foolproof, contrary to mainstream opinion. In the transition from simple data to advanced data, biometric frameworks might be better able to analyze advanced data. With a fingerprint reader, fingerprints can be lifted from an ordinary object, even though it is difficult to accomplish. It is possible to create fingerprints under gum or putty in addition to making copies. In addition, they explore voice and facial recognition, where duplicates and photographs can be made to trick biometric systems. Integrating biometric units with Active Directory (AD), LDAP is a practical method for protecting and securing the verification information. LDAP and AD offer encrypted storage for biometric accreditations, as well as credentials such as client IDs and passwords. 

In the health facility, there was a conspiracy among employees, which affected operations, including providing services to patients who were not entitled to certain health benefits. Biometrics will curtail such practices. We recommended separating roles for key parties involved in the approval and credentials processes, performing high inspections, maintaining strong firewalls, and defining access levels. In addition, all confidential records are kept on a central database server at the hospital. As a result, hackers can cause huge damage if they gain access to the database by exposing millions of users to security threats. Providing customers with the ability to store their data on their devices is essential to Acacia medical center. Additionally, 'salting' techniques can be used to create features that are not linkable in order to guarantee differentiation from other variants. This is achieved by adding a randomly generated implementation of a hashing algorithm to scramble the data. 

\subsection{Future Work}
In this study, we identified and discussed issues and challenges related to implementing biometrics in the healthcare sector in Ghana.
Performing an extensive stakeholder analysis in this area would be a worthwhile research project in the future, since there are varying perspectives on cost and benefit analyses for biometrics technology. Furthermore, we suggest that it would be beneficial to conduct an extensive study to examine the needs and requirements of the various actors (patients, healthcare professionals, and healthcare providers) who are the stakeholders in the usage of biometric systems.
\bibliographystyle{ACM-Reference-Format}
\bibliography{references}

\end{document}